\documentclass[pre,twocolumn,amsmath,amssymb,nofootinbib,floatfix,superscriptaddress]{revtex4}

\usepackage{enumerate}
\usepackage{graphicx,bm}
\usepackage[normalem]{ulem}

\usepackage{graphicx,bm,xcolor}
\usepackage{enumerate}
\usepackage{graphicx,bm}
\usepackage[normalem]{ulem}

\newcommand{\rev}[1]{\textcolor{blue}{#1}}

\begin{document}

\title{Strong-coupling critical behavior in three-dimensional lattice 
  Abelian gauge models with charged
  $N$-component scalar fields and $SO(N)$ symmetry}

\author{Claudio Bonati} 
\affiliation{Dipartimento di Fisica dell'Universit\`a di Pisa
        and INFN Sezione di Pisa, Largo Pontecorvo 3, I-56127 Pisa, Italy}

\author{Andrea Pelissetto}
\affiliation{Dipartimento di Fisica dell'Universit\`a di Roma Sapienza
        and INFN Sezione di Roma, I-00185 Roma, Italy}

\author{Ettore Vicari} 
\affiliation{Dipartimento di Fisica dell'Universit\`a di Pisa,
        Largo Pontecorvo 3, I-56127 Pisa, Italy}

\date{\today}

\begin{abstract}
  We consider a three-dimensional lattice Abelian Higgs gauge model
  for a charged $N$-component scalar field ${\bm \phi}$, which is
  invariant under $SO(N)$ global transformations for generic values of
  the parameters.  We focus on the strong-coupling regime, in which
  the kinetic Hamiltonian term for the gauge field is a small
  perturbation, which is irrelevant for the critical behavior.  
  The Hamiltonian depends on a parameter $v$ which
  determines the global symmetry of the model and the symmetry
  of the low-temperature phases. 
  We present renormalization-group predictions, based on a 
  Landau-Ginzburg-Wilson effective description that relies 
  on the identification of the appropriate order parameter and  on the 
  symmetry-breaking patterns that occur at the strong-coupling 
  phase transitions.
  For $v=0$, the global symmetry group of the model is $SU(N)$;
  the corresponding model may undergo continuous transitions only for $N=2$. For
  $v\not=0$, i.e., in the $SO(N)$ symmetric case, continuous transitions (in the
  Heisenberg universality class) are possible also for $N=3$ and 4.
  We perform Monte Carlo simulations for $N=2,3,4,6$, to verify 
  the renormalization-group predictions. Finite-size scaling analyses of 
  the numerical  data are in full agreement.
\end{abstract}

\maketitle


\section{Introduction}
\label{intro}

Lattice Abelian Higgs (AH) models, in which an Abelian gauge field
interacts with a charged $N$-component degenerate scalar field ${\bm
  \phi}$, provide an effective description of many collective
phenomena characterized by the interplay of topological gauge
excitations and scalar fluctuations~\cite{Anderson-book,Wen-book}. In
particular, they provide examples of topological transitions that are
not characterized by the breaking of a global symmetry
\cite{Herbut-book,Fradkin-book,Sachdev-19,Fradkin-23,SBSVF-04}.  The
phase diagram of this class of systems has been extensively studied,
see, e.g., Refs.~\cite{HLM-74,FS-79,DH-81,CC-82,
  KK-85,KK-86,BN-86,BN-86-b,BN-87,MU-90,KKS-94,BFLLW-96,
  HT-96,KKLP-98,OT-98,CN-99,HS-00,KNS-02,MHS-02,
  SSSNH-02,SSNHS-03,NRR-03,NSSS-04,SSS-04,
  WBJSS-05,CFIS-05,TIM-05,CIS-06,KPST-06, WBJS-08,MV-08,HBBS-13,
  BS-13,FH-17,FH-19,PV-19-CP,PV-19-AH3d,PV-20-mfcp,
  PV-20-largeNCP,BPV-20-hcAH,BPV-21-ncAH,BPV-21-bgs,WB-21, BPV-22-mpf,
  BPV-22,BPV-23b,BPV-23c,BPV-24}, characterizing the different phases
in terms of the topological properties of the gauge correlations, and
identifying the possible symmetry-breaking patterns.

The global symmetry of the model and the symmetry breaking that occurs
at phase transitions depend on the scalar self-interactions. Most of
the investigations considered $SU(N)$-symmetric scalar potentials.
The phase diagrams and the critical behaviors that occur in this class of
models have been extensively investigated in the literature, see e.g.,
Refs.~\cite{HLM-74,FS-79,SSS-04,FH-17,PV-19-CP,
  PV-19-AH3d,PV-20-largeNCP,BPV-20-hcAH,BPV-21-ncAH,WB-21,BPV-22,BPV-23b,
  BPV-24}.  However, as discussed in Refs.~\cite{MU-90,BPV-23c}, one
may also consider more complex scalar self-interactions, which are
invariant under a smaller group of transformations, which preserves
some irreducible permutation of the field components, to avoid transitions in
which only some of the components become critical (in this case the
effective theory would 
be of interest for the analysis of the
multicritical behavior).  By considering more general scalar
potentials and different global symmetry groups, one is able to
determine the variety of critical behaviors that can be observed in
the presence of an emergent Abelian gauge symmetry in generic lattice
systems.

In this work, we consider the two-parameter quartic scalar potential
\begin{equation}
  V_O ({\bm \phi})= r \,\bar{\bm\phi} \cdot {\bm\phi}  + u
  \,(\bar{\bm\phi} \cdot {\bm\phi})^2 + v \,|{\bm\phi}\cdot
    {\bm\phi}|^2.
  \label{vopot}
\end{equation}
For $v = 0$ the potential is $SU(N)$ symmetric, while for $v\not=0$ it
is only invariant under $O(N)$ transformations.  Results for this
model were presented in Ref.~\cite{BPV-23c}.  Here we extend, and
verify numerically, the renormalization-group (RG) predictions 
for the phase transitions that occur in the gauge strong-coupling limit.

We consider a three-dimensional (3D) lattice $U(1)$ gauge model,
obtained by a straightforward discretization of the AH field theory
\begin{eqnarray}
{\cal L} = \frac{1}{4 g^2} \sum_{\mu\nu} F_{\mu\nu}^2 + 
   \sum_{\mu} |D_\mu{\bm\phi}|^2 + V_O({\bm \phi}).
  \label{OAHFT}
\end{eqnarray}
We observe in passing that this gauge field theory can also be derived
starting from an $O(2)\otimes O(N)$ invariant real scalar model, by
gauging the $O(2)$ global group~\cite{BPV-23c}.  To simplify the
model, we consider the limit $r\to -\infty$ and $u\to\infty$
keeping $r/u=-2$ fixed, which forces $\phi$ to be a unit vector.
Thus, in the lattice model we 
associate an $N$-component unit-length complex vector ${\bm z}_{\bm
  x}$ (satisfying $\bar{\bm z}_{\bm x} \cdot {\bm z}_{\bm x} =1$) with
each site of a cubic lattice.  Concerning the gauge field, one can
consider compact formulations, in which the fundamental field is a
complex phase $\lambda_{{\bm x},\mu}$, or noncompact formulations, in
which the basic gauge variable is $A_{{\bm x},\mu}\in {\mathbb R}$ and
$\lambda_{{\bm x},\mu}$ is defined as $\lambda_{{\bm
    x},\mu}=e^{iA_{{\bm x},\mu}}$. In both cases the Hamiltonian
reads~\cite{BPV-23c}
\begin{eqnarray}
&&H = H_z + \kappa \, K_g, \label{LAH}\\ &&H_z= - 2NJ \sum_{{\bm
      x},\mu} {\rm Re}\,( \lambda_{{\bm x},\mu} \bar{\bm z}_{\bm x}
  \cdot {\bm z}_{{\bm x}+\hat\mu}) + v \,\sum_{\bm x}|{\bm z}_{\bm
    x}\cdot {\bm z}_{\bm x}|^2, \nonumber
\end{eqnarray}
where $\kappa\sim g^{-2}$ is the inverse gauge coupling, and $K_g$ is
the gauge-field Hamiltonian term, which assumes different forms
in compact and noncompact formulations.

We focus on the strong-coupling regime $\kappa/J\ll 1$, in which the gauge
kinetic term $\kappa\,K_g$ gives only rise to a small irrelevant perturbation.
Therefore, to study the strong-coupling critical behavior, we do not need to
specify the form of $K_g$.  Actually, we can limit our analyses to the model
(\ref{LAH}) with $\kappa = 0$, neglecting the gauge term $\kappa\,K_g$, because
the critical behavior for finite (sufficiently small) values of $\kappa$ is
expected to be the same as along the $\kappa=0$ line, as discussed below.  As a
consequence of the irrelevance of the gauge kinetic term, the critical behavior
in the strong-coupling regime can be determined by considering effective
Landau-Ginzburg-Wilson (LGW) theories in terms of gauge-invariant
composite scalar operators only. The only role of the gauge degrees of
freedom in the present model is thus that of preventing some correlators (the
nongauge-invariant ones) from becoming critical, forcing us to consider 
a gauge-invariant order parameter.

\begin{figure}[tbp]
\includegraphics*[width=0.95\columnwidth]{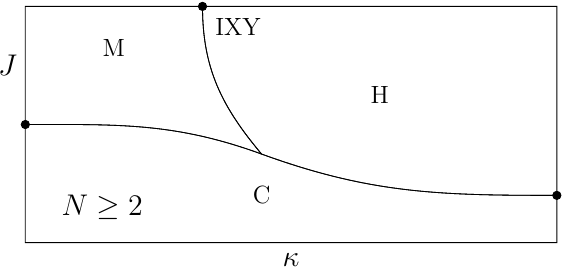}
  \caption{The $\kappa$-$J$ phase diagram of the $SO(N)$ lattice AH
    model with noncompact gauge fields, for $N\ge 2$ and generic
    values of $v$.  Three phases are present: the small-$J$ Coulomb
    (C) phase, in which the scalar field is disordered and gauge
    correlations are long ranged; the large-$J$ molecular (M) and
    Higgs (H) ordered phases, in which the global symmetry is
    spontaneously broken. The results we present in this work refer to
    the strong-coupling CM line that starts at $\kappa = 0$.  }
\label{phadiaLAH}
\end{figure}

For $N\ge 2$, the phase diagram of the noncompact AH model (\ref{LAH}) presents
two different low-temperature (large-$J$) phases, in which the global symmetry
is spontaneously broken and that differ in the topological properties of the
gauge correlations.  The symmetry breaking pattern depends on the number $N$ of
components and on the Hamiltonian parameter $v$~\cite{BPV-23c}.  A sketch of
the $\kappa$-$J$ phase diagram for the noncompact AH model is shown in
Fig.~\ref{phadiaLAH}, for $N\ge 2$ and generic values of $v$.  The $\kappa$-$J$
phase diagram of the corresponding compact models differ substantially for
sufficiently large values of $\kappa$, see, e.g.,
Refs.~\cite{PV-19-AH3d,BPV-20-hcAH,BPV-21-ncAH,BPV-22}. However, the phase
diagrams are qualitatively the same in the strong-coupling regime.  Indeed, the
behavior for $\kappa/J\ll 1$ is the same as for $\kappa=0$
and the nature of the gauge fields is irrelevant in the latter case. Thus, to determine the
critical behavior along the Coulomb-Molecular transition line reported in
Fig.~\ref{phadiaLAH} (noncompact formulation) or along the analogous line that
occurs in compact models it is enough to consider the case $\kappa = 0$.

Beside the Abelian $U(1)$ gauge invariance, the lattice model
(\ref{LAH}) has a global $SO(N)$ symmetry, ${\bm \phi} \to S {\bm
  \phi}$ with $S\in SO(N)$, which enlarges to $SU(N)$ for $v=0$. 
The global
symmetry is broken at a finite-temperature disorder-order transition,
whose nature depends on $N$ and on the sign of the Hamiltonian
parameter $v$~\cite{BPV-23c}.

For $\kappa = 0$ and $v = 0$, the lattice model (\ref{LAH}) reduces to
the $SU(N)$ symmetric CP$^{N-1}$ model. For $N=2$, it can be mapped
onto an $O(3)$-vector model, and thus it shows a continuous transition
in the Heisenberg universality class. For any $N\ge 3$ the transition
is of first order~\cite{PV-19-CP}.  The nature of the transitions
changes for $v \ne 0$, as a consequence of the smaller global symmetry
of the model.  As we shall see, in the presence of $SO(N)$ invariance,
continuous transitions also occur for $N=3$ and $N=4$, for positive
values of the scalar self-interaction parameter $v$.

In this paper we report a numerical study of the model (\ref{LAH})
with $\kappa = 0$. We perform Monte Carlo (MC) simulations for several
values of $N$ and determine the nature of the critical transitions
using finite-size scaling (FSS) methods.  The results nicely support
the predictions obtained by using an effective LGW description of the
system in terms of properly defined gauge-invariant order parameters.
For $N=2$, the $O(3)$-vector continuous transition at $v=0$ turns into
two continuous transition lines for $v\not=0$. They belong to the
Ising and $XY$ universality class for $v>0$ and $v<0$, respectively.
For $v=0$ and any $N\ge 3$ transitions are of first order.  Only
first-order transitions are also expected for any $N\ge 5$ in the
$SO(N)$ invariant model. However, for $N=3$ and $N=4$ it is possible
to observe continuous transitions for $v> v^* > 0$, where $v^*$ is
positive and corresponds to a tricritical point.  For $v < v^*$
transitions are of first order. The continuous transitions belong to
the $O(3)$ vector universality class for both values of $N$, but the 
underlying mechanism is different. For $N=3$ the Heisenberg
behavior is a consequence of the fact that the order parameter is
equivalent to a three-component real vector.  For $N=4$ the effective
description involves two three-component real vectors, and the $O(3)$
behavior follows from a nonperturbative RG analysis that shows that
the interaction between these two fields is irrelevant in the critical
limit.

The paper is organized as follows. In Sec.~\ref{sec2} we present the
theoretical analysis of the model. In Sec.~\ref{sec2.1} we summarize
the general results obtained in Ref~\cite{BPV-23c}, while in
Sec.~\ref{sec2.2} we present a field-theoretical analysis of the
effective LGW model appropriate to describe the $SO(N)$ AH model for
$v>0$.  In Sec.~\ref{sec3} we present our numerical results that
confirm the theoretical predictions. Conclusions are presented in
Sec.~\ref{sec4}.  The Appendix presents some technical field-theory
results that are relevant for $N\ge 4$.

\section{Effective LGW description of the transitions} \label{sec2}

\subsection{General arguments} \label{sec2.1}

Let us now review the main results on the critical behavior of the
model in the strong-coupling regime obtained in
Ref.~\cite{BPV-23c}. The critical behavior along the strong-coupling
transition line that starts at $\kappa = 0$ depends on the sign of the
parameter $v$, which determines the symmetry breaking pattern.  The
symmetry of the low-temperature phases can be determined by analyzing
the minima of the scalar potential (\ref{vopot}).  For $v > 0$, the
fields corresponding to the minimum configurations can be parametrized
as \cite{BPV-23c}
\begin{equation}
  \bm{\phi} = {1\over \sqrt{2}}\left( \bm{s}_1 + i \bm{s}_2\right), \qquad
  \bm{s}_1 \cdot \bm{s}_2 = 0,
\label{minimum2}
\end{equation}
where $\bm{s}_1$ and $\bm{s}_2$ are orthogonal real vectors satisfying
$|\bm{s}_1| = |\bm{s}_2|$. In this case the global
$SO(N)$ symmetry of the model is broken to $SO(2)\oplus O(N-2)$.

For $v < 0$, the minimum configurations can be parametrized as 
\begin{equation}  
  \bm\phi = e^{i\alpha} \bm{s}, 
\label{minimum1}
\end{equation}
where $\bm{s}$ is a real $N$-component vector, and $\alpha$ an
arbitrary phase. The $SO(N)$ symmetry is broken to $O(N-1)$.

To characterize the spontaneous breaking of the $SO(N)$ symmetry, two
different order parameters were introduced,
\begin{eqnarray}
 R_{L,\bm x}^{ab} &=& {1\over 2}(\bar{z}_{\bm
    x}^a z_{\bm x}^b + \bar{z}_{\bm x}^b z_{\bm x}^a) - {1\over
    N}\delta^{ab},
  \label{wqoper} \\ 
  T_{L,\bm x}^{ab} &=& {1\over 2i} (\bar{z}_{\bm x}^a
  z_{\bm x}^b - \bar{z}_{\bm x}^b z_{\bm x}^a) ,
  \label{wqopera} 
\end{eqnarray}
which transform under two different representations of the $SO(N)$
group.  Their behavior depends on the sign of $v$.  For $v < 0$,
$R_{L,\bm x}^{ab}$ condenses in the ordered phase, while $T_{L,\bm
  x}^{ab}$ vanishes.  For $v > 0$ and $N =2$, $T_{L,\bm x}^{ab}$
condenses, while $R_{L,\bm x}^{ab}$ vanishes. Finally, for $v > 0$ and
$N \ge 3$, both order parameters condense in the ordered phase.

For sufficiently small values of $\kappa$ along the CM transition line
(or along the corresponding line in compact models), gauge
fluctuations are not expected to play an active role at the
transition. Indeed, the gauge properties of the two small-$\kappa$
phases are the same: gauge modes are long ranged and charged
excitations are confined in both of them.  Therefore, the transition
should be uniquely driven by the breaking of the global
symmetry. Thus, an effective description of the critical universal
behavior can be obtained by considering a LGW theory for an
appropriate gauge-invariant scalar order parameter that condenses at
the transition, without considering the gauge fields
~\cite{PV-19-CP,PV-19-AH3d,BPV-21-ncAH}.

For $v < 0$ the relevant order parameter \cite{BPV-23c} is $R_{L,\bm
  x}^{ab}$. The antisymmetric operator $T_{L,\bm x}^{ab}$ is expected
to be disordered on both sides of the transition. Since $R_{L,\bm
  x}^{ab}$ is a real symmetric operator, we expect the small-$\kappa$
transitions to be described by a LGW for a real symmetric traceless
$N\times N$ matrix field $\Phi^{ab}({\bm x})$, that represents a
coarse-grained average of $R_{L,\bm x}^{ab}$ over a large, but finite,
lattice domain.  The corresponding LGW Lagrangian is obtained by
considering all monomials in $\Phi^{ab}({\bm x})$ that are allowed by
the global $SO(N)$ symmetry up to fourth order.  We obtain
\begin{eqnarray}
  {\cal L}_\Phi &=& {\rm Tr} (\partial_\mu \Phi)^2 + r \,{\rm Tr} \,\Phi^2
  + s \,{\rm tr} \,\Phi^3
\label{hlg}   \\
&+&   \,u \, ({\rm Tr}
\,\Phi^2)^2 + v\, {\rm Tr}\, \Phi^4.
\nonumber
\end{eqnarray}
For $N=2$, we can parametrize the field as 
\begin{equation}
  \Phi = \begin{pmatrix} 
        \phi_1 & \phi_2 \\
        \phi_2 & -\phi_1
       \end{pmatrix}  \; .
\end{equation}
It follows that $(\Phi^2)^{ab} = (\phi_1^2 + \phi_2^2) \delta^{ab}$,
the cubic term vanishes, and the two quartic terms are equivalent.
The resulting LGW theory is equivalent to that of the $O(2)$-symmetric
vector model. Thus, we predict continuous transitions to belong to the
$XY$ universality class.  On the other hand, for $N\ge 3$ the cubic
$\Phi^3$ term is generally present. This is usually considered as the
indication that phase transitions are of first order, as one can
easily infer using mean-field arguments.  We expect this behavior to
hold for any $v<0$, up to $v=0$, where we recover the
$SU(N)$-invariant $CP^{N-1}$ model, whose transition is continuous for
$N=2$, in the $O(3)$ vector universality class, and of first order for
any $N\ge 3$ \cite{PV-19-CP,PV-20-largeNCP}.

As discussed in Ref.~\cite{BPV-23c}, for $v>0$ the relevant order
parameter is the antisymmetric tensor field $T^{ab}_{L,\bm x}$. We
shall therefore consider the LGW model for an antisymmetric $N\times
N$ real field $\Psi^{ab}({\bm x})$, which represents the
coarse-grained average of $T_{L,\bm x}^{ab}$.  The corresponding LGW
Lagrangian reads
\begin{eqnarray}
  {\cal L}_\Psi &=&  {\rm Tr} \, \partial_\mu \Psi^t \partial_\mu \Psi
  + r \,{\rm Tr} \,\Psi^t\Psi\nonumber\\
&+&   \,u \, ({\rm Tr}
\,\Psi^t\Psi)^2 + w\, {\rm Tr}\, (\Psi^t\Psi)^2,
\label{hlgt} 
\end{eqnarray}
where $\Psi^t=-\Psi$ is the transpose of $\Psi$.  Note that the cubic
term is absent because ${\rm Tr} \Psi^n = 0$ for any odd $n$.  As
discussed in Ref.~\cite{BPV-23c}, also the operator $R_{L,\bm x}^{ab}$
is expected to be critical at transitions with $v>0$ for any $N\ge 3$.
The analysis of the behavior for $v\to \infty$ shows that in this
limit we have the relation
\begin{equation}
  R^{ab}_L = - a \left[(T^2_L)^{ab} - {\delta^{ab}\over N}  
    \hbox{Tr } T^2_L \right],
\label{R-T2}
\end{equation}
where $a$ is a positive constant. In the LGW formalism, this implies
that $R^{ab}_L$ has the same critical behavior as
\begin{equation}
   {\cal R}^{ab} = (\Psi^2)^{ab} - {\delta^{ab}\over N} \hbox{Tr} \Psi^2.
\label{defcalR}
\end{equation}
This relation should hold for any continuous transition with $v > 0$.

For $N=2$ and $N=3$ the LGW Lagrangian (\ref{hlgt}) can be simplified
\cite{AKL-13,BPV-23c}.  For $N = 2$ we can write $\Psi^{ab}$ in terms
of a single real scalar field $\phi$ defined by $\Psi^{ab} =
\epsilon^{ab}\phi$. The two quartic terms are equivalent, and we
obtain the LGW model for a real scalar field. Continuous transitions
are therefore expected to belong to the Ising universality class. Note
that ${\cal R}^{ab} = 0$ in this case, which implies that $R^{ab}_L$
is not critical for $N=2$.

For $N=3$ we can write $\Psi^{ab}(x)$ in terms of a single
three-component vector as $\Psi^{ab} = \epsilon^{abc}\phi^c$, where
$\epsilon^{abc}$ is the completely antisymmetric tensor. Again, the
quartic terms are equivalent and we obtain the $O(3)$ vector LGW
Hamiltonian.  Thus, continuous transitions should belong to the $O(3)$
vector universality class. As for the operator ${\cal R}^{ab}$, we obtain 
\begin{equation}
{\cal R}^{ab} = \phi^a \phi^b - {1\over 3} \delta^{ab} \phi^2.
\end{equation}
This relation implies that $R^{ab}_L$ should have the same critical
behavior as the spin-two operator in the Heisenberg model.

No simplifications occur for $N\ge 4$. To determine the critical
behavior one should therefore study the RG flow of the model
(\ref{hlgt}) in the space of the quartic couplings $u$ and $w$. As
discussed in the Appendix, two different types of symmetry breakings
are possible in model (\ref{hlgt}), depending on the sign of $w$. An
ordered phase with $SO(2)\oplus O(N-2)$ symmetry is obtained for $w <
0$. Therefore, continuous transitions for $v > 0$ are only possible if
the LGW field theory admits a stable fixed point with $w < 0$.

\subsection{Field-theory analysis of the effective LGW model for $v < 0$}
\label{sec2.2}

In this Section we perform a field-theory analysis of the RG flow in the model
with Lagrangian (\ref{hlgt}) for $N\ge 4$, with the purpose of studying the
possible existence of stable RG fixed points with $w < 0$.  For this purpose we
consider the $\epsilon$-expansion approach. The $\beta$ functions have been
computed to three-loop order in Refs.~\cite{AKL-13, AKL-17}. At two-loop 
order  we have
\begin{eqnarray}
&& \beta_u(u,w) = 
\textstyle
     -\epsilon u + 
     {1\over 12} (N^2 - N + 16) u^2  + 
     {1\over 4} w^2  \nonumber \\ [0.5mm]
&& \qquad \textstyle
     + {1\over 6} (2 N - 1) u w 
    - {1\over 24} (3 N^2 - 3 N + 28) u^3  \nonumber \\[0.5mm]
&& \qquad \textstyle
    - {11\over 36} (2 N - 1) u^2 w 
    - {1\over 288} (5 N^2 - 5 N + 164)  u w^2 \nonumber \\[0.5mm]
&& \qquad \textstyle
    - {1\over 48} (2 N - 1) w^3,\\[0.5mm]
&& \beta_w(u,w) = 
\textstyle
     - \epsilon w + 2 u w  + {1\over 12} (2 N - 1) w^2
\nonumber \\[0.5mm]
&& \qquad \textstyle
   - {1\over 72} (5 N^2 - 5 N + 164)  u^2 w 
   - {11\over 36} (2 N - 1) u w^2 \nonumber \\[0.5mm]
&& \qquad  \textstyle
   - {1\over 96} (N^2 - N + 20) w^3.
\end{eqnarray}
At one loop, beside the trivial fixed
point $u=w=0$, the $\beta$ functions always have a zero on the $w=0$
axis. This fixed point corresponds to an $O(K)$ invariant
[where $K=N(N-1)/2$] theory and is always unstable. Indeed, the $w$ term is
a spin-four perturbation of the fixed point, which is always relevant
for $N\ge 4$ \cite{CPV-02,HV-11,Chester-etal-20-o3}.  Two additional
fixed points are present, but only for relatively small values of $N$;
more precisely, for
\begin{equation}
N < N^*(\epsilon) \approx  {1\over 4} (2 + 3 \sqrt{22}) - 
   {9\epsilon \over 16 \sqrt{22} } \approx 4.018 - 0.120 \epsilon,
\end{equation}
with corrections of order $\epsilon^2$.  Given the small negative
correction term, it seems plausible to assume that $N^* < 5$ in three
dimensions ($\epsilon=1$), which implies that no stable fixed points
exist for $N\ge 5$. We thus predict transitions to be of first order
for any $N\ge 5$.

Let us now discuss the model with $N=4$. In this case the
antisymmetric tensor $\Psi^{ab}$ transforms under a reducible
representation of the $SO(4)$ group.  It is therefore convenient to
parametrize $\Psi^{ab}$ in terms of two three-component vectors
$\phi_1^e$ and $\phi_2^e$ ($e=1,2,3$) that transform irreducibly:
\begin{equation}
\Psi^{ef} = {1\over 2} \sum_g \epsilon^{efg} (\phi_1^g - \phi_2^g),
\quad \Psi^{4f} = {1\over 2} (\phi_1^f + \phi_2^f),
\end{equation}
for $e,f,g=1,2,3$.  In terms of these two fields we obtain the
Lagrangian
\begin{eqnarray}
  {\cal L}_\phi&=& {1\over 2} \sum_{i=1}^2
  [(\partial_\mu \phi_i)^2 + r \phi_i^2 ] 
\nonumber \\
&& + \Bigl(u + {3\over 4} w\Bigr) (\phi_1^2 + \phi_2^2)^2 - 
     {w\over 2} (\phi_1^4 + \phi_2^4).
\end{eqnarray}
This model is known in the literature as $MN$ model
\cite{Aharony-73,GL-76,Aharony-76,Shpot-88,PV-02,PV-05} and represents
the most general model in which $M$ $N$-component real vector
fields ($M=2$ and $N=3$ in our case) interact symmetrically.

Beside the unstable fixed point with $w = 0$, the model admits a
second simple fixed point that corresponds to two noninteracting
$O(3)$ vector fields. Indeed, since for $u + 3w/4 = 0$ the two vector
fields decouple, there is a fixed point with
\begin{equation}
   u = {3\over 2} U^*_{O(3)}  \qquad 
   w = - 2 U^*_{O(3)}, 
\end{equation}
where $U^*_{O(3)} > 0$ is the fixed point of the $O(3)$ Lagrangian
\begin{equation}
L_{O(3)} = {1\over 2} (\partial_\mu\varphi)^2 + {r\over 2} \varphi^2 + 
     U (\varphi^2 )^2.
\end{equation}
It is easy to prove nonperturbatively that this fixed point is
stable. Indeed, the RG dimension of the perturbation is $y_p =
2/\nu_{O(3)} - d = \alpha_{O(3)}/\nu_{O(3)}$, as it corresponds to an
energy-energy interaction between the two scalar fields.  Since
$\alpha_{O(3)} < 0$ in the $O(3)$ model, the interaction is irrelevant
and thus the fixed point is stable.  The fixed point lies in the
region $w < 0$ and is therefore relevant for the model with $v >
0$. Thus, we predict that continuous transitions for $N=4$ belong to
the $O(3)$ universality class. Note, however, that $y_p$ is very small,
$y_p \approx -0.19$, and thus we expect slowly decaying scaling
corrections to the critical behavior.

To determine the critical behavior of ${\cal R}^{ab}$ defined in
Eq.~(\ref{defcalR}), we express it in terms of $\phi_1$ and
$\phi_2$. We obtain
\begin{eqnarray}
{\cal R}^{ef} &=& -{1\over 2} (\phi_1^e \phi_2^f + \phi_1^f \phi_2^e) + 
    {1\over 2} \delta^{ef} \phi_1\cdot \phi_2, \\
{\cal R}^{44} &=& - {1\over 2}  \phi_1\cdot \phi_2, \qquad
{\cal R}^{4e} = - {1\over 2}  \sum_{fg} \epsilon^{efg} \phi_1^f \phi_2^g ,
\nonumber 
\end{eqnarray}
where $e$, $f$, $g$ run from 1 to 3. These relations show that
$R_L^{ab}$ behaves as the product of two independent $O(3)$ vector
fields.

\subsection{Summary} \label{sec2.3}

The previous analysis and the results of Ref.~\cite{BPV-23c} allow us
to predict the behavior of the model in the strong coupling regime
$\kappa \ll 1$. For $N=2$ we expect Ising transitions for $v > 0$ and
$XY$ transitions for $v < 0$. The line with $v = 0$ is a multicritical
line where the symmetry group enlarges to $O(3)$ and we observe the same
critical behavior as in the $CP^1$ model.

For $N=3$ and $N=4$, we expect first-order transitions for $v < 0$ (no
stable fixed points exist in the LGW effective theory) and also for
$v=0$, as in the $CP^{N-1}$ model \cite{PV-19-CP}.  For $v > 0$
continuous transitions are possible, belonging to the $O(3)$
universality class in both cases (but with slowly decaying scaling
corrections for $N=4$). Since, the transition is 
of first order for $v=0$, i.e., in the CP$^{N-1}$ model~\cite{PV-19-CP}, it is
natural to expect first-order transitions also for small positive
values of $v$. As a consequence, we predict the existence of a
tricritical positive value $v^*$, such that the transition is in the
Heisenberg universality class for $v > v^*$ and of first order for 
$v < v^*$.

Finally, for $N\ge 5$ no stable fixed points occur in the LGW RG flow
and thus we expect transitions to be of first order in all cases.

\section{Numerical results} \label{sec3}

In this section we present numerical Monte Carlo (MC) results, with
the purpose of verifying the predictions of the previous Section.  We
consider the model with $\kappa = 0$ and partition function
\begin{equation}
Z = \sum_{\{z,\lambda\}} e^{-H_z({\bm z},\lambda)},
\label{zdef}
\end{equation}
(we set $\beta=1/T=1$) and perform several runs by varying $J$ around the
critical point for $N=2,3,4$, and 6. We consider cubic lattices of size $L^3$
with periodic boundary conditions and use a combination of Metropolis and, for
the gauge field $\lambda$, microcanonical updates.\footnote{For $\bm z$ we use
Metropolis updates with two different proposals: a) we select two components
$i,j$ and perform a real rotation, ${z'}_i =z_i \cos \alpha + z_j \sin\alpha$,
${z'}_i = -z_i \sin \alpha + z_j \cos\alpha$; b) we select a single component
and propose ${z'}_i = e^{i\alpha} z_i$. For $\lambda_\mu$, we consider a
Metropolis update with $\lambda_\mu' = e^{i\alpha} \lambda_\mu$. In all cases
$\alpha$ is chosen in an interval $[-\theta,\theta]$, where $\theta$ 
guarantees an acceptance of approximately
40\% (different values of $\theta$ are used in the three cases above). For
$\lambda$ we also use a microcanonical update. If $F = z_{\bm x} \cdot
\bar{z}_{\bm x + \mu}$, we perform the update $\lambda_{{\bm x},\mu}' =
\bar{\lambda}_{{\bm x},\mu}
F/\overline{F}$. }

\subsection{Observables and finite-size scaling relations}

To characterize the critical behavior we consider correlations of the
order parameters. We consider the two-point correlation function of
the operator $R_L^{ab}$,
\begin{eqnarray}
G_R({\bm x}-{\bm y}) &= &
   \sum_{ab}
   \langle R^{ab}_{L,\bm x} R^{ba}_{L,\bm y} \rangle,
\label{gT}
\end{eqnarray}
and the analogous quantity $G_T({\bm x}-{\bm y})$ for $T_L^{ab}$. 
Then, we define the Fourier transform
\begin{equation}
\widetilde{G}_\#({\bm p}) =
    {1\over V} \sum_{{\bm x}-{\bm y}}
      e^{i{\bm p}\cdot ({\bm x} - {\bm y})} G_\#({\bm x},{\bm y})
\end{equation}
($V$ is the volume) of the two correlation functions.  The
corresponding susceptibilities and correlation lengths are defined as
\begin{eqnarray}
&&\chi_\# = 
\widetilde{G}_\#({\bm 0}),
\label{chisusc}\\
&&\xi^2_\# \equiv  {1\over 4 \sin^2 (\pi/L)}
{\widetilde{G}_\#({\bm 0}) - \widetilde{G}_\#({\bm p}_m)\over
\widetilde{G}_\#({\bm p}_m)},
\label{xidefpb}
\end{eqnarray}
where ${\bm p}_m = (2\pi/L,0,0)$.

In our FSS analysis we use RG invariant quantities. We consider
\begin{equation}
R_{\xi,\#} = \xi_\#/L
\end{equation}
and the Binder parameters. We define $B_R$ as 
\begin{equation}
B_R = {\langle \mu_{2,R}^2\rangle \over \langle \mu_{2,R}\rangle^2} , \qquad
\mu_{2,R} = \sum_{{\bm x}{\bm y}} 
\sum_{ab}
   R^{ab}_{L,\bm x} R^{ba}_{L,\bm y}.
\label{binderdef}
\end{equation}
The definition of $B_T$ is analogous.

For $N=4$, we also consider the operators 
\begin{equation}
\phi_\pm^A = T_{L}^{A4} \pm {1\over 2} \sum_{BC} \epsilon^{ABC} T_{L}^{BC},
\end{equation}
where all indices run from 1 to 3. As already discussed, these two quantities
transform irreducibly under SO(4) rotations. The correlation functions 
\begin{equation}
     G_{\phi,\pm} ({\bm x} - {\bm y})
    = \sum_A \langle \phi_{\pm,\bm x}^A \phi_{\pm,\bm y}^A  \rangle
\end{equation}
satisfy $G_{\phi,+}({\bm x}) = G_{\phi,-}({\bm x})$ and 
$G_{T}({\bm x})= - G_{\phi,+}({\bm x})- G_{\phi,-}({\bm x})$. In particular,
the correlation length computed using $G_{\phi,\pm}({\bm x})$ is the same as 
$\xi_T$. The Binder parameter is instead different. We define 
\begin{equation}
B_\phi = 
{1\over 2} {\langle \mu_{2,+}^2\rangle \over \langle \mu_{2,+}\rangle^2} + 
{1\over 2} {\langle \mu_{2,-}^2\rangle \over \langle \mu_{2,-}\rangle^2} 
\qquad
\mu_{2,\pm} = \sum_{{\bm x}{\bm y}} 
\sum_A
   \phi^{A}_{\pm,\bm x} \phi^{A}_{\pm,\bm y}.
\end{equation}
At continuous transitions, in the FSS limit, the Binder parameter as
well as any renormalization-group invariant quantity $R$ scales as
\begin{equation}
R(J,L) \approx f_R(X) + L^{-\omega} f_{c,R}(X), \quad X = (J-J_c)
L^{1/\nu} ,
\label{rsca}
\end{equation}
where $\omega$ is the leading 
correction-to-scaling exponent, and $J_c$ gives the
position of the critical point.
Relation (\ref{rsca}) can also be written as 
\begin{equation}
R(\beta,L) = F_R(R_\xi) + L^{-\omega} F_{c,R}(R_\xi) + \ldots
\label{r12sca}
\end{equation}
where $F_R(x)$ is universal---it only depends on the universality class, the 
boundary conditions, and the lattice shape---and 
$F_{c,R}(x)$ is universal apart from a
multiplicative constant. Relation (\ref{r12sca}) will play an important role 
to identify the universality class: To verify that the models belong 
to the Ising, XY, and Heisenberg universality classes, as predicted above, 
we will compare the curves $F_R(R_\xi)$ computed in the present model with 
those computed in the corresponding $N$-vector model with the same boundary
conditions. If the identification is correct, the data we obtain here should 
converge towards the corresponding $N$-vector curves as $L$ increases.

Critical exponents can also be obtained from the FSS analysis. The
exponent $\nu$ can be obtained by fitting the data to
Eq.~(\ref{rsca}). The exponent $\eta$ instead can be obtained by
fitting the susceptibility data to
\begin{equation}
\chi = L^{2-\eta} [G_\chi (X) + O(L^{-\omega})],
\end{equation}
where $X$ is defined in Eq.~(\ref{rsca}). Numerically, however, it is
more convenient to fit the data to
\begin{equation}
\chi = L^{2-\eta} [\tilde{G}_\chi (R_\xi) + O(L^{-\omega})],
\label{chisca}
\end{equation}
since these fits do not require any knowledge of $\nu$ and $J_c$.

\subsection{Strong-coupling critical behavior for $N = 2$}

\begin{figure}[tbp]
\includegraphics*[width=0.95\columnwidth]{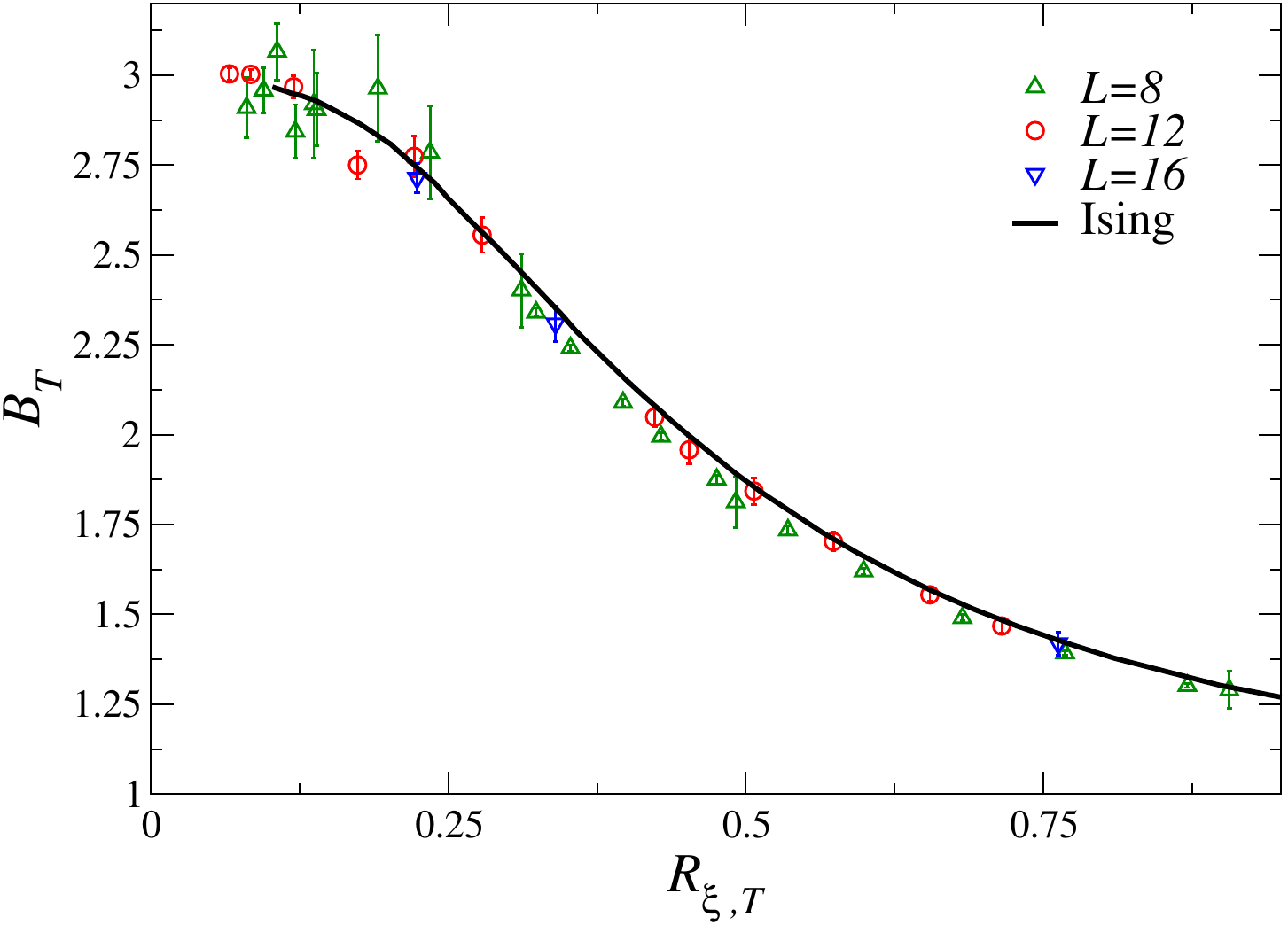}
\includegraphics*[width=0.95\columnwidth]{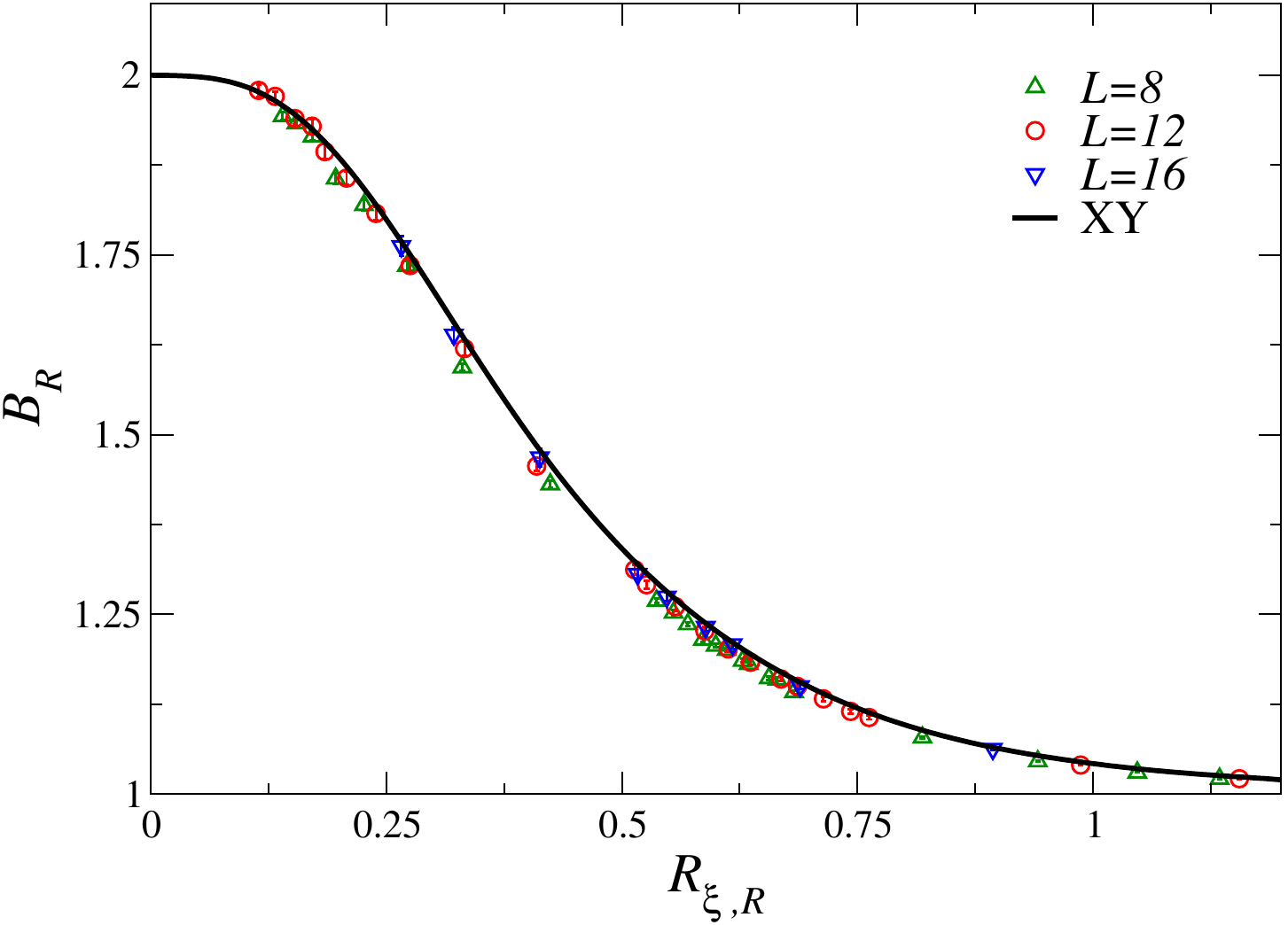}
  \caption{Top: Plot of $B_T$ versus $R_{\xi,T}$ for $v = 10$; Bottom:
    Plot of $B_R$ versus $R_{\xi,R}$ for $v = -10$.  In both cases
    $N=2$ and $\kappa = 0$. The continuous curves have been computed
    in the Ising model (upper panel) and in the $XY$ model (lower
    panel). The relative error on the curves is approximately of
    0.5\%.}
\label{fig:B-xiT-N2}
\end{figure}

To determine the critical behavior for $N=2$, we have performed MC
simulations at $\kappa=0$, varying $J$.  We have only considered
relatively small lattice sizes ($L\le 16$), as the results 
already confirm quite precisely the predictions of the
previous Section.

First, we set $v = 10$. We observe a critical transition for $J
\approx 0.37$, which we expect to be an Ising transition. To verify
it, in the upper panel of Fig.~\ref{fig:B-xiT-N2} we report the Binder
parameter $B_T$ versus $R_{\xi,T}$ and compare the data with the curve
computed in the Ising model. We observe good scaling, in spite of the
fact that lattices are quite small. To further confirm the
predictions, we fit $B_T$ and $R_{\xi,T}$ to
Eq.~(\ref{rsca}). Parametrizing the universal curve with a polynomial,
we obtain $\nu=0.61(3)$ and $\nu=0.64(2)$ from the analysis of $B_T$
and $R_{\xi,T}$, respectively, in good agreement with the Ising result
\cite{KPSV-16} $\nu_I=0.629971(4)$. Finally, to determine $J_c$
precisely, we repeat the fits fixing $\nu$ to the Ising value,
obtaining $J_c = 0.3741(5)$.

An analogous analysis has been performed for $v = -10$. In the lower
panel of Fig.~\ref{fig:B-xiT-N2} we report the Binder parameter $B_R$
versus $R_{\xi,R}$ and compare the data with the curve computed in the
$XY$ model. Again, we observe good agreement confirming the LGW
prediction. To estimate $J_c$ we have fitted the two RG invariant
ratios to Eq.~(\ref{rsca}), fixing $\nu=\nu_{XY} =0.6717(1)$
\cite{CHPV-06,Hasenbusch-19,CLLPSSV-20}.  We obtain $J_c = 0.5633(3)$.

\subsection{Strong-coupling critical behavior for $N = 3$}

\begin{figure}[tbp]
\includegraphics*[width=0.95\columnwidth]{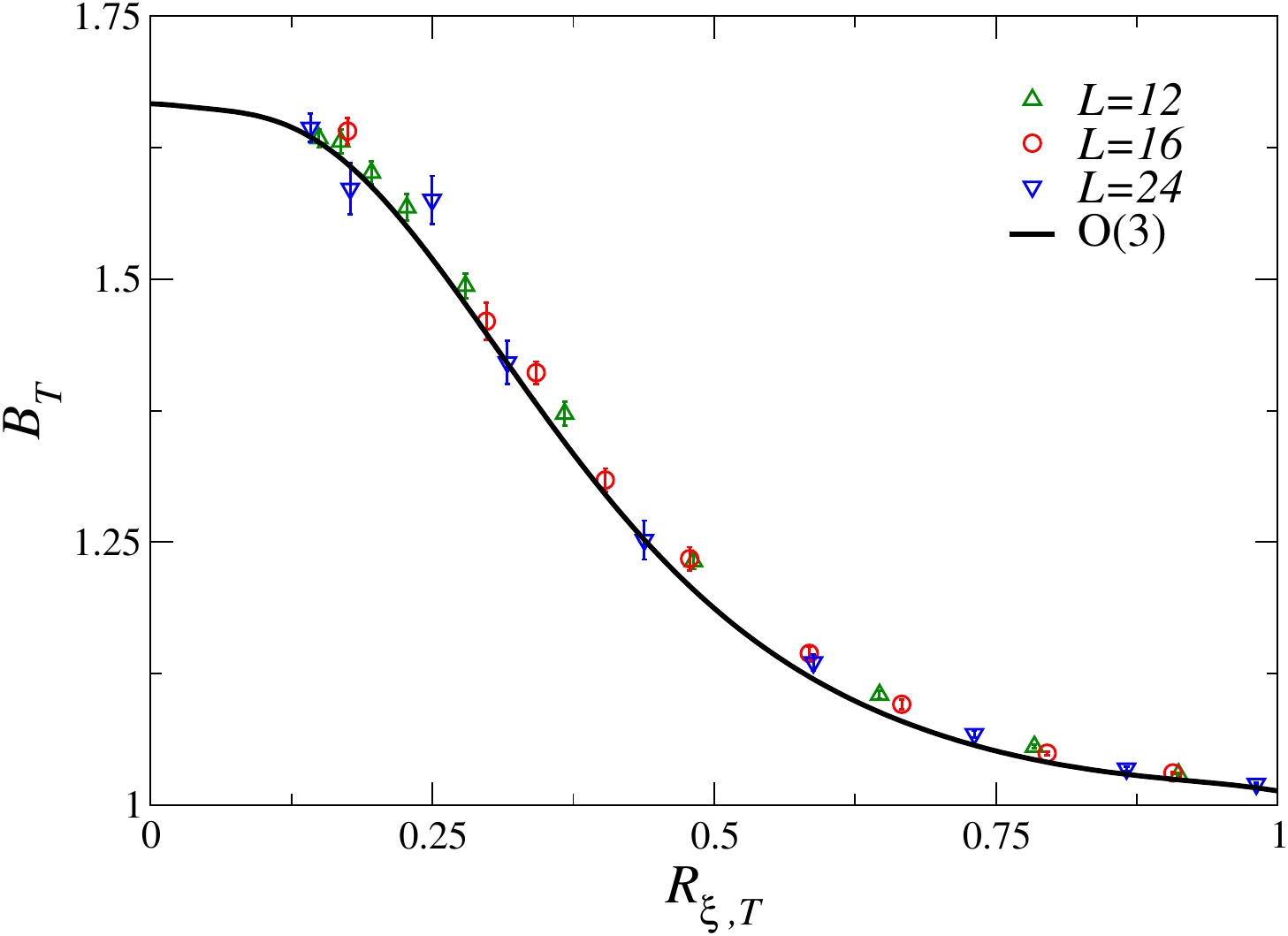}
\includegraphics*[width=0.95\columnwidth]{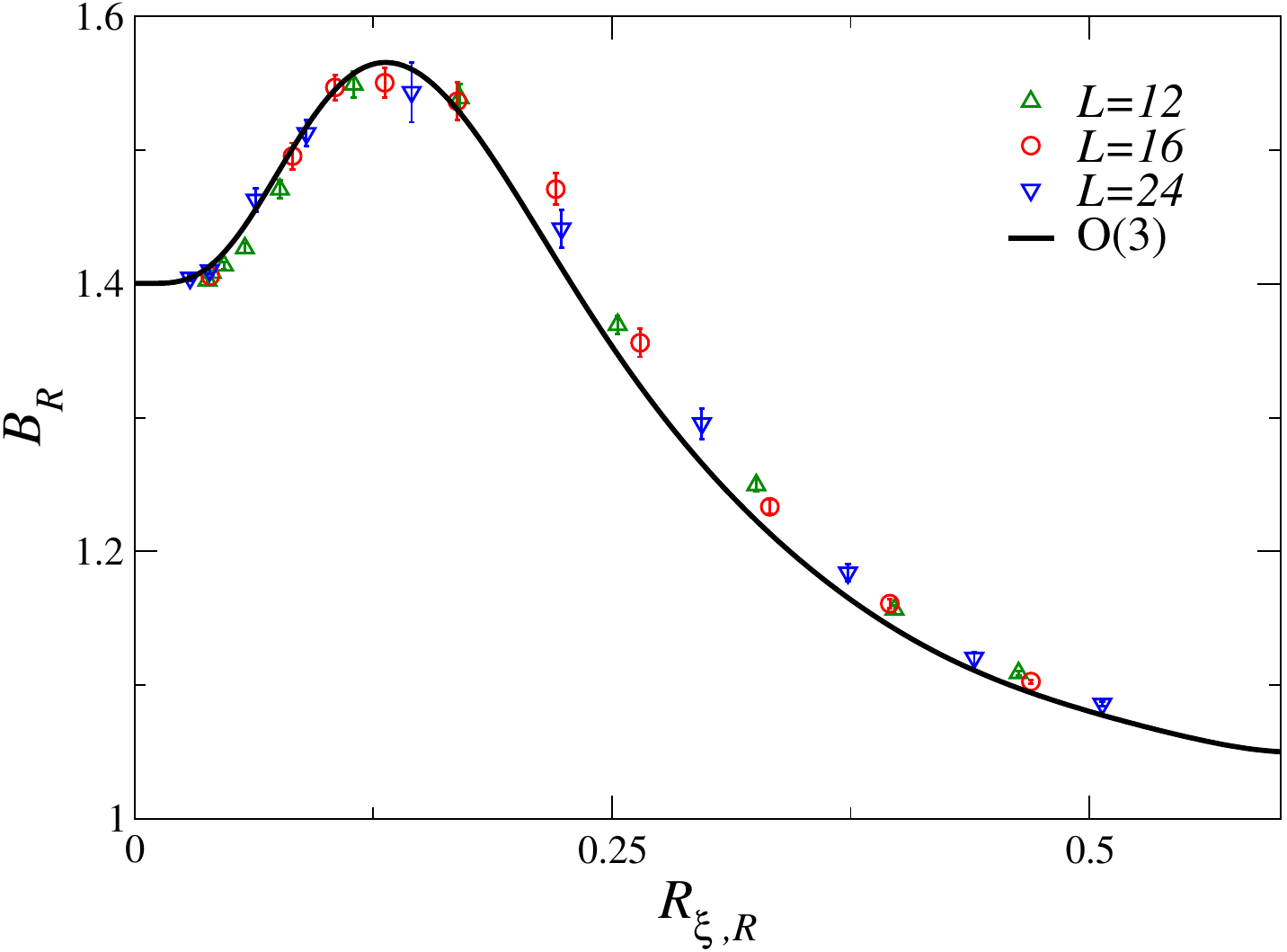}
\caption{Top: plot of $B_T$ versus $R_{\xi,T}$ for different values of
  $L$; Bottom: plot of $B_R$ versus $R_{\xi,R}$. Data for $\kappa =
  0$, $v = 10$, and $N=3$. The continuous curves have been computed in
  the Heisenberg $O(3)$ vector model. In the top panel we report the
  curve for vector (spin-1) observables; in the lower panel we report
  the curve for tensor (spin-2) observables (the relative error on
  these curves is approximately 0.5\%).}
\label{fig:URxi-N3}
\end{figure}

For $N=3$ we have performed a numerical analysis for $v=10$ and
$\kappa =0$ to verify the predicted behavior. A priori, the transition
is expected to be either of first order (this occurs if $v < v^*$,
where $v^*$ is the tricritical point), or continuous in the Heisenberg
universality class.  The numerical results are consistent with an
$O(3)$ continuous transition.  Indeed, if we plot the Binder parameter
$B_T$ versus $R_{\xi,T}$, the results fall quite precisely on the
corresponding universal curve for vector correlations in the
Heisenberg model, see the upper panel of Fig.~\ref{fig:URxi-N3}. As an
additional check, we have fitted the estimates of $B_T$ and
$R_{\xi,T}$ to Eq.~(\ref{rsca}), obtaining $\nu = 0.73(2)$, which is
consistent with the accurate estimate~\cite{Hasenbusch-20} $\nu =
0.71164(10)$ for the Heisenberg universality class, see also
Refs.~\cite{KPSV-16,KP-17,CHPRV-02}. To determine $J_c$, we have
repeated the fits fixing $\nu$ to the $O(3)$ value, obtaining $J_c =
0.4479(3)$.

As we discussed in Sec.~\ref{sec2.1}, the correlations of the 
field $R_{L,\bm x}$ should behave as the correlations of the 
spin-two operator (it is defined as 
$\Sigma^{ab} = \sigma^a \sigma^b - \delta^{ab}/3$, 
where $\sigma^a$ is the 3-component Heisenberg spin) in the 
$O(3)$ model. To verify this prediction, in the lower panel of 
Fig.~\ref{fig:URxi-N3} we report $B_R$ versus $\xi_R/L$, together with 
the Heisenberg scaling curve for $B_\Sigma$ versus $\xi_\Sigma/L$,
where the latter quantities are computed from correlations of the 
spin-two operator $\Sigma^{ab}$. We observe a reasonable agreement. 
Tiny deviations are observed for intermediate values of $R_{\xi,R}$,
presumably the result of corrections to scaling.

\subsection{Strong-coupling critical behavior for $N = 4$}

\begin{figure}[tbp]
\includegraphics*[width=0.95\columnwidth]{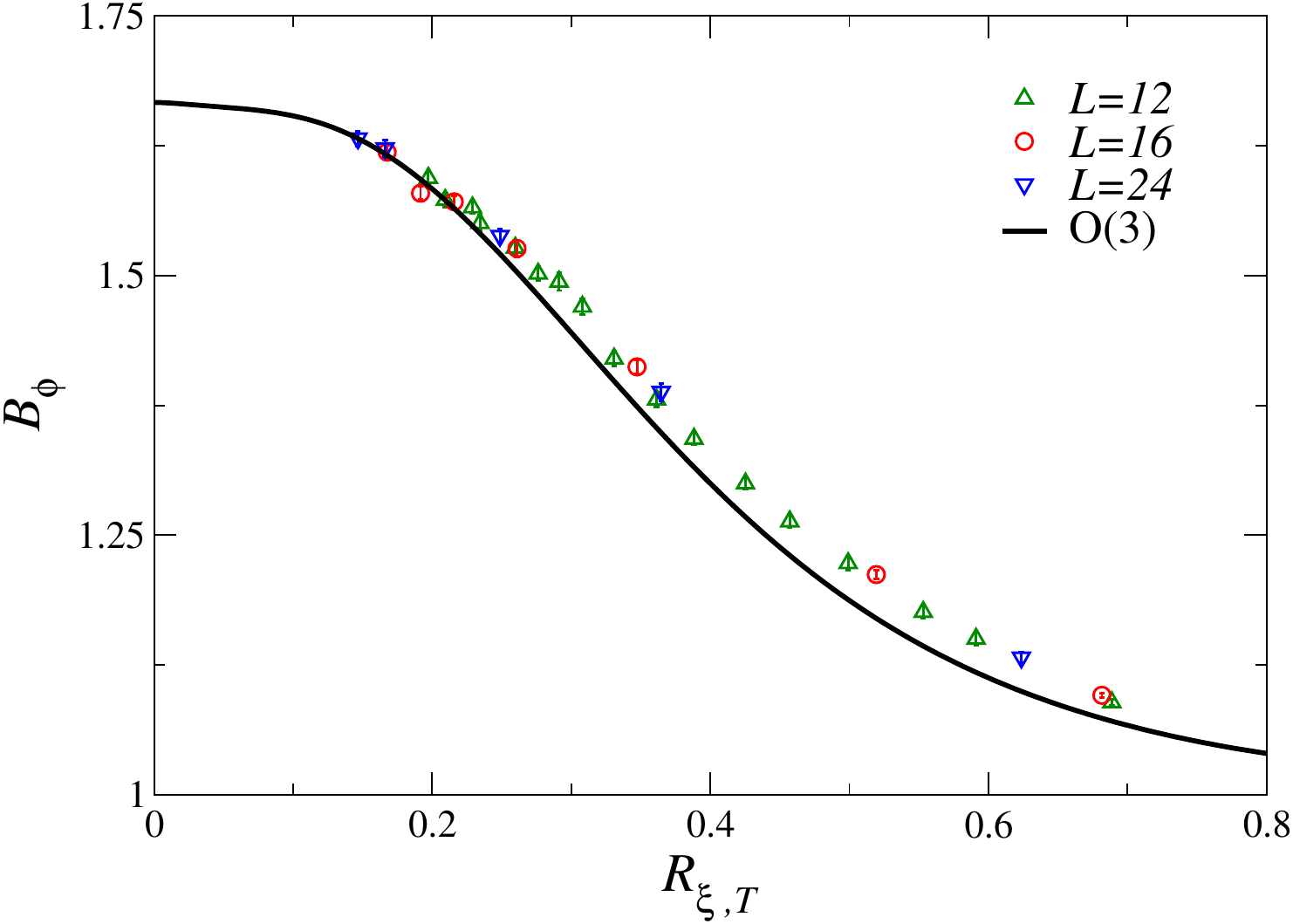}
\includegraphics*[width=0.95\columnwidth]{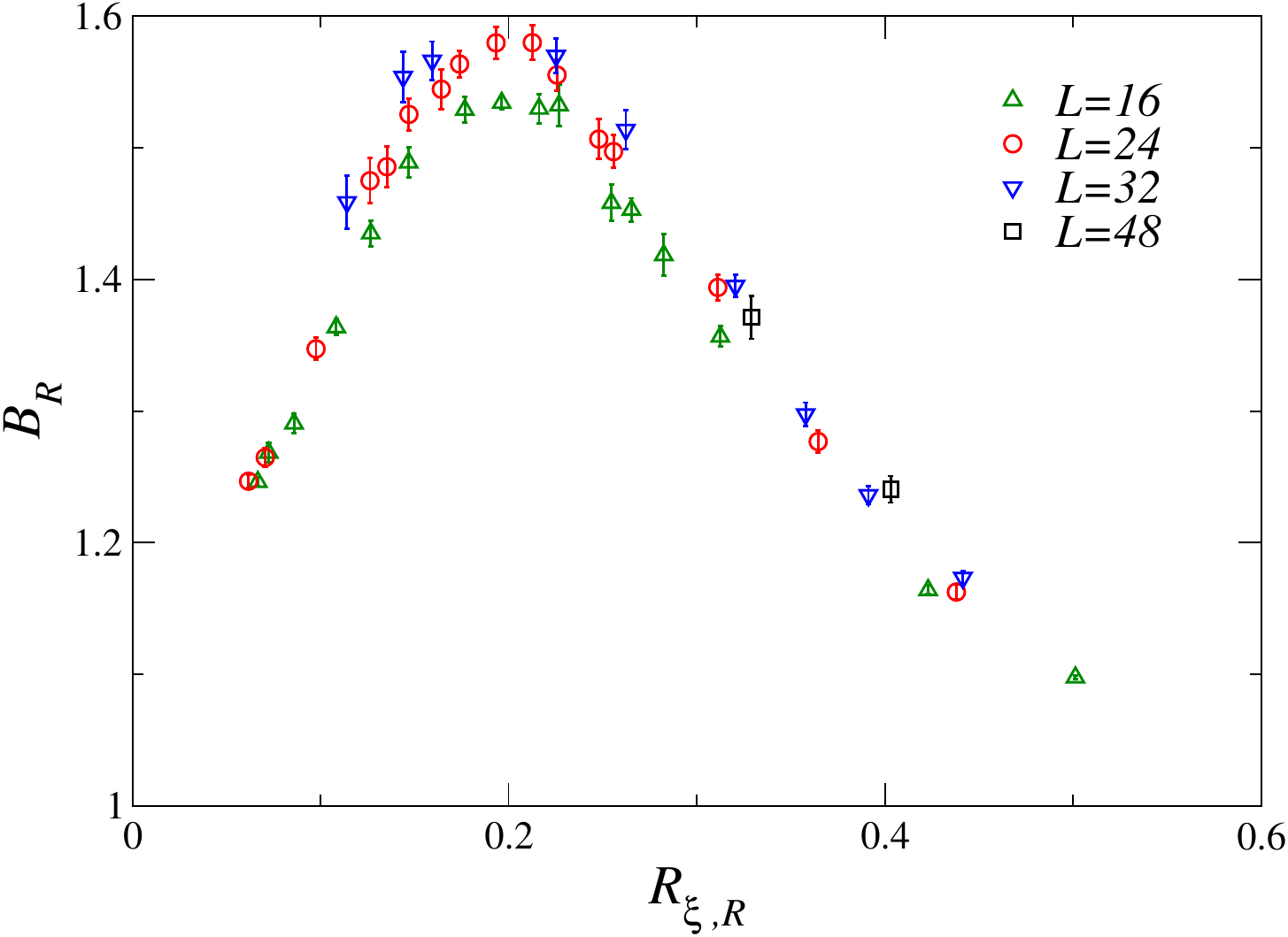}
\caption{Plot of $B_\phi$ versus $R_{\xi,T}$ (top) and of $B_R$ versus
  $R_{\xi,R}$ (bottom).  Data for $\kappa = 0$, $v = 10$, and
  $N=4$. The continuous curve in the upper panel has been computed in
  the Heisenberg $O(3)$ model, using vector (spin-1) correlations.
  The relative error on the curve is approximately 0.5\%.  }
\label{fig:URxi-N4}
\end{figure}

\begin{figure}[tbp]
\includegraphics*[width=0.95\columnwidth]{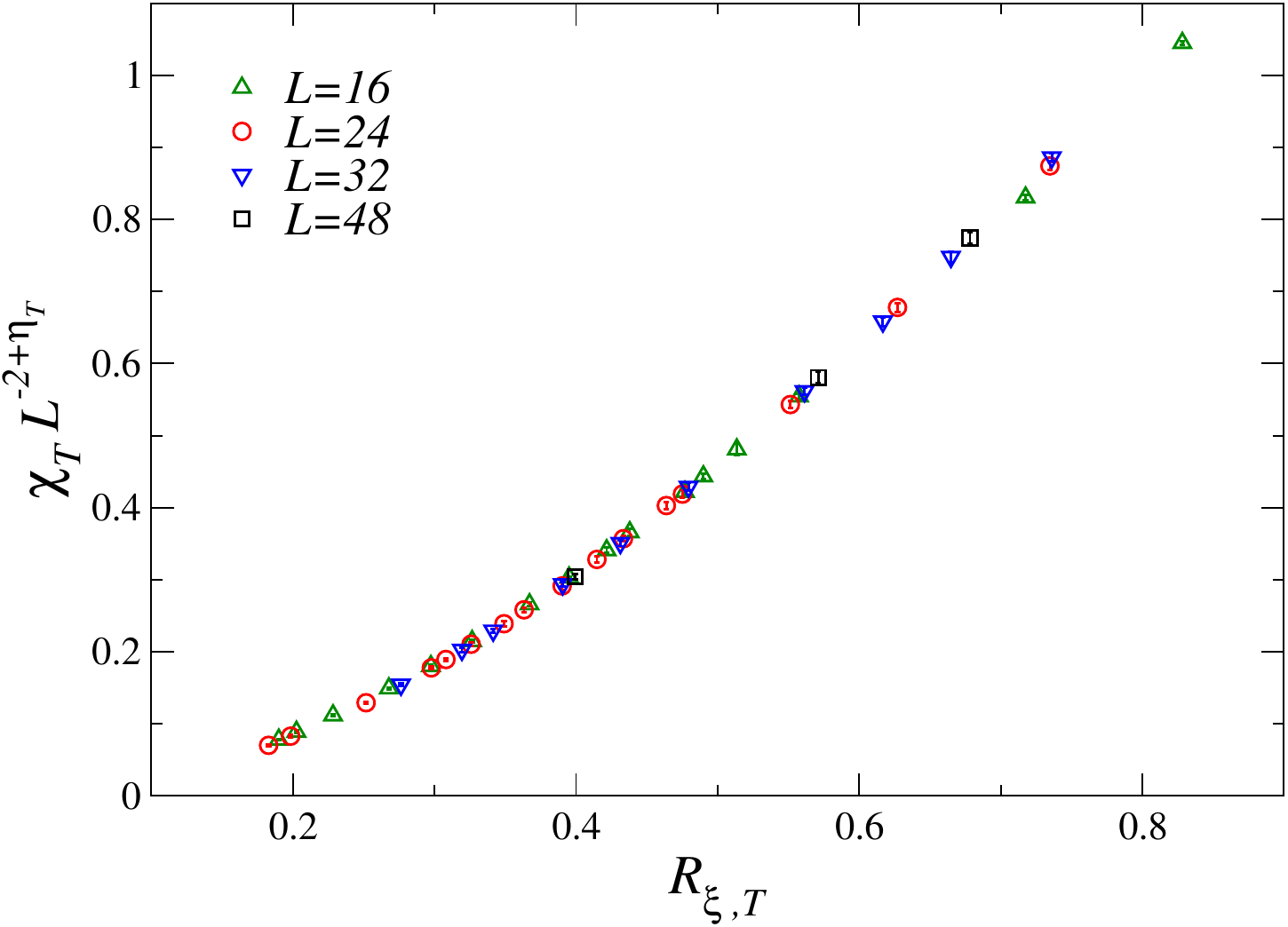}
\includegraphics*[width=0.95\columnwidth]{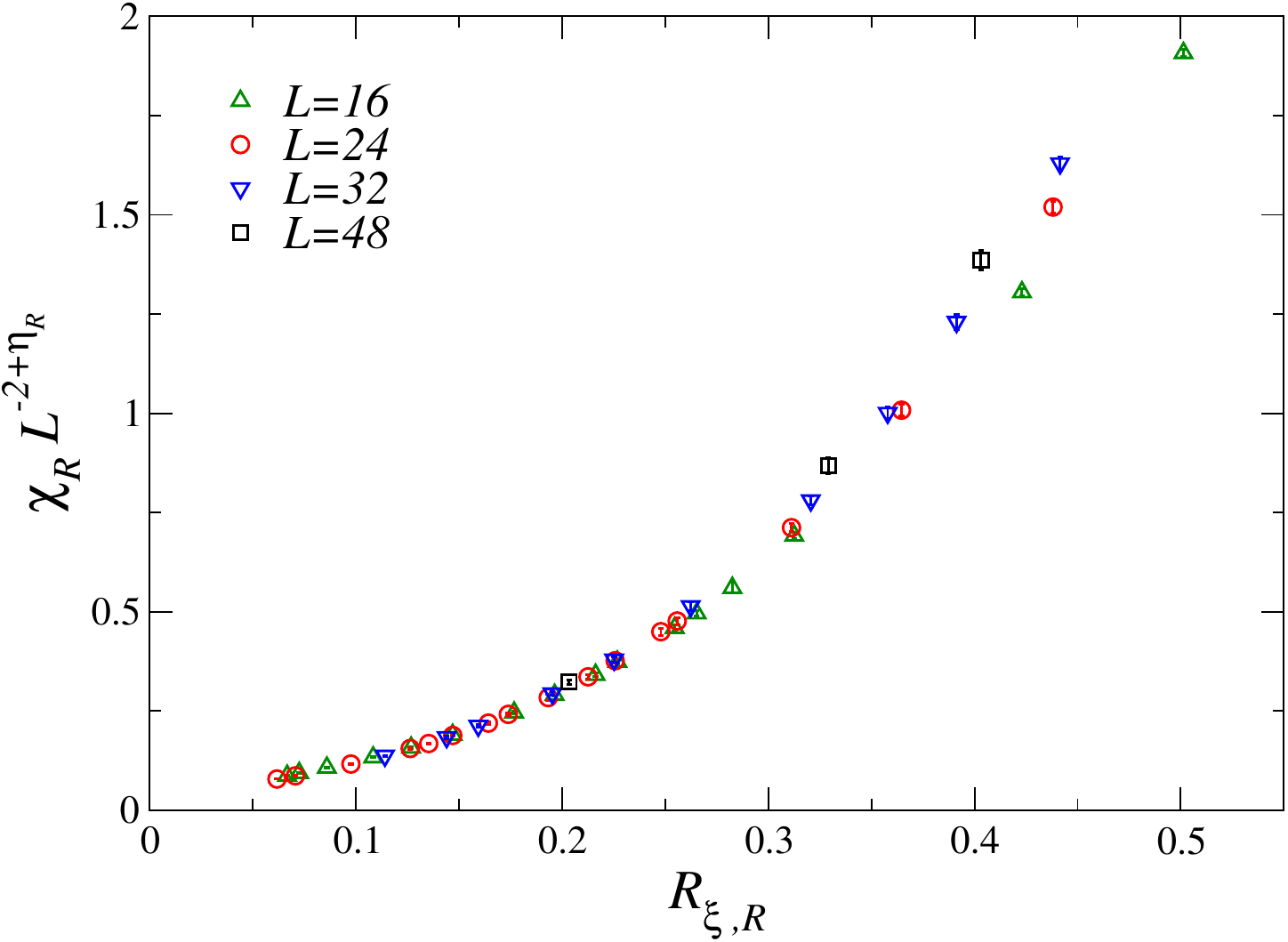}
\caption{Plot of $L^{-2+\eta_T} \chi_T$ versus $R_{\xi,T}$ (top) and
  of $L^{-2+\eta_R} \chi_R$ versus $R_{\xi,R}$ (bottom).  Data for
  $\kappa = 0$, $v = 10$, and $N=4$.  We set $\eta_T = \eta_H$ and
  $\eta_R = 1 + 2 \eta_H$, where $\eta_H$ is the vector susceptibility
  exponent in the Heisenberg O(3) model: $\eta_H = 0.0362$. }
\label{fig:Uchi-N4}
\end{figure}

For $N=4$ we have investigated the critical behavior for $v=10$ and
$\kappa =0$. If we plot the Binder parameters $B_R$, $B_T$, and
$B_\phi$ versus the $R_{\xi,R}$ and $R_{\xi,T}$ we observe good
scaling, indicating that the transition is continuous, see
Fig.~\ref{fig:URxi-N4}. To verify the arguments of Sec.~\ref{sec2.2}
and, in particular, whether the critical behavior belongs to the
Heisenberg universality class, we compare the plot of $B_\phi$ versus
$R_{\xi,T}$ with the corresponding curve computed in the Heisenberg
model, see the upper panel of Fig.~\ref{fig:URxi-N4}.  The numerical
data are close to the Heisenberg curve, although some systematic
deviations are clearly visible, especially for intermediate values of
$R_{\xi,T}$, i.e., close to the critical point. These small deviations
can be easily explained by the presence of slowly decaying scaling
corrections due to the $\phi_1^2\phi_2^2$ in the LGW approach. They
decay very slowly, as $L^{-0.19}$, making it very difficult to observe
the asymptotic behavior.  For instance, to reduce scaling corrections
by a factor of two, one should increase the lattice size be a factor
of 38, which is is clearly not feasible.

To provide additional evidence for the correctness of the LGW
predictions, we consider the susceptibilities $\chi_R$ and
$\chi_T$. The susceptibility $\chi_T$ should scale as the magnetic
susceptibility in the Heisenberg model.  Therefore, data should scale
as in Eq.~(\ref{chisca}) with
\cite{Hasenbusch-20,KPSV-16,KP-17,CHPRV-02} $\eta_T = \eta_H =
0.0362(1)$. This prediction is verified in
Fig.~\ref{fig:Uchi-N4}. Data scale very well, as predicted.  A second
important consistency check is provided by the analysis of
$\chi_R$. The arguments of Sec.~\ref{sec2.2} indicate that $R_L$
behaves as the product of two independent $O(3)$ vector fields.  This
implies that $G_T({\bm x})$ has the same critical behavior as
$G_{H}({\bm x})^2$, where $G_{H}({\bm x}) = \langle \sigma_{\bm 0}
\cdot \sigma_{\bm x}\rangle$ is the vector correlation function in the
Heisenberg model ($\sigma$ is the fundamental variable in the
Heisenberg model). At the critical point, $G_{H}({\bm x})$ scales as
$|x|^{-1-\eta_H}$. Therefore, we have
\begin{equation}
\chi_T \sim \int d^3 x\,  G_{H}({\bm x})^2 \sim 
    \int^L r^2 dr\, r^{-2-2 \eta_H} \sim L^{1-2\eta_H}.
\end{equation}
It follows that $\chi_R$ scales as in Eq.~(\ref{chisca}) with $\eta_R
= 1 + 2 \eta_H = 1.0724(1)$. This prediction is tested in
Fig.~\ref{fig:Uchi-N4}. Again, data scale quite well, confirming the
LGW predictions.

\subsection{Strong-coupling critical behavior for $N = 6$}

\begin{figure}[tbp]
\includegraphics*[width=0.95\columnwidth]{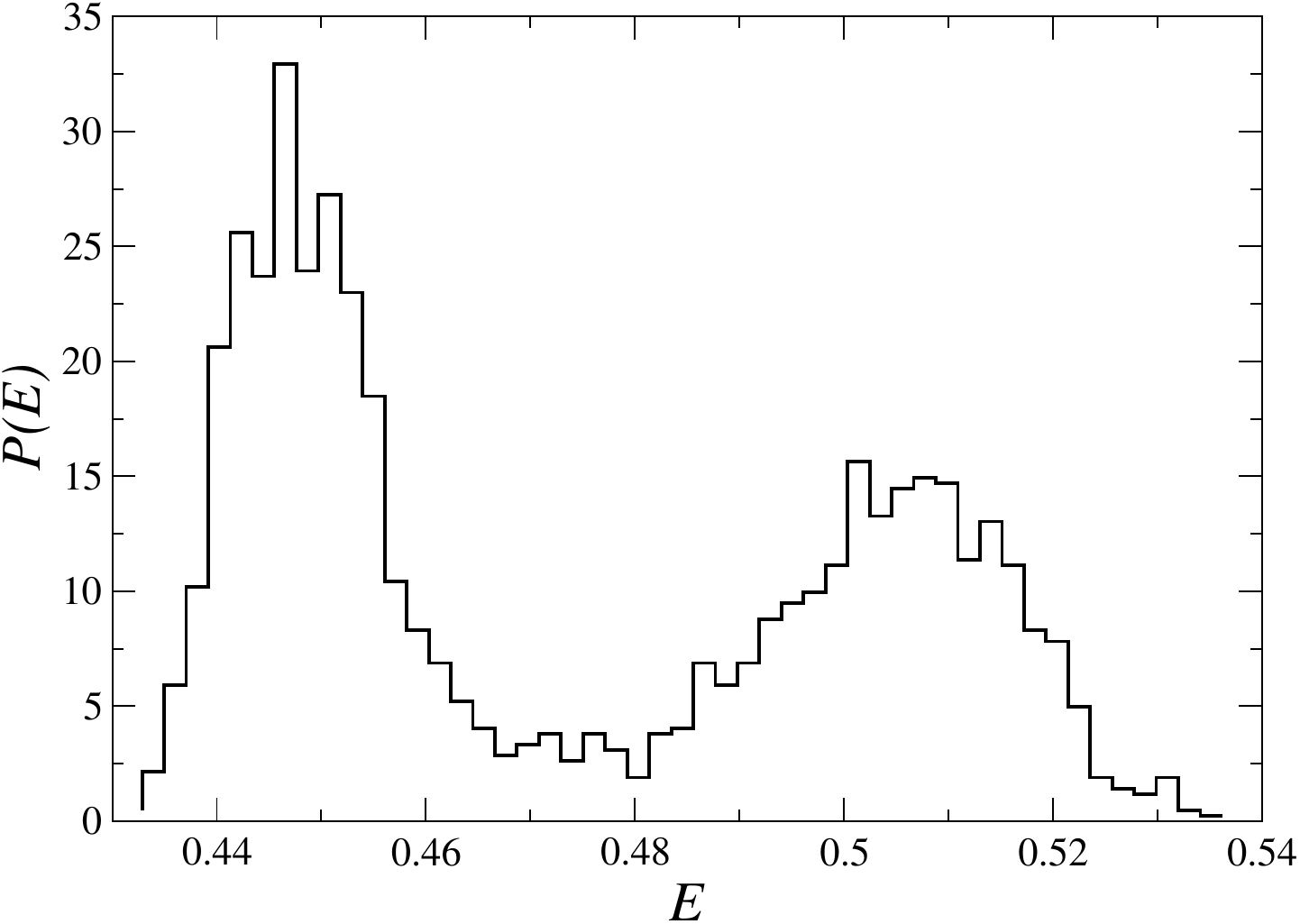}
\includegraphics*[width=0.95\columnwidth]{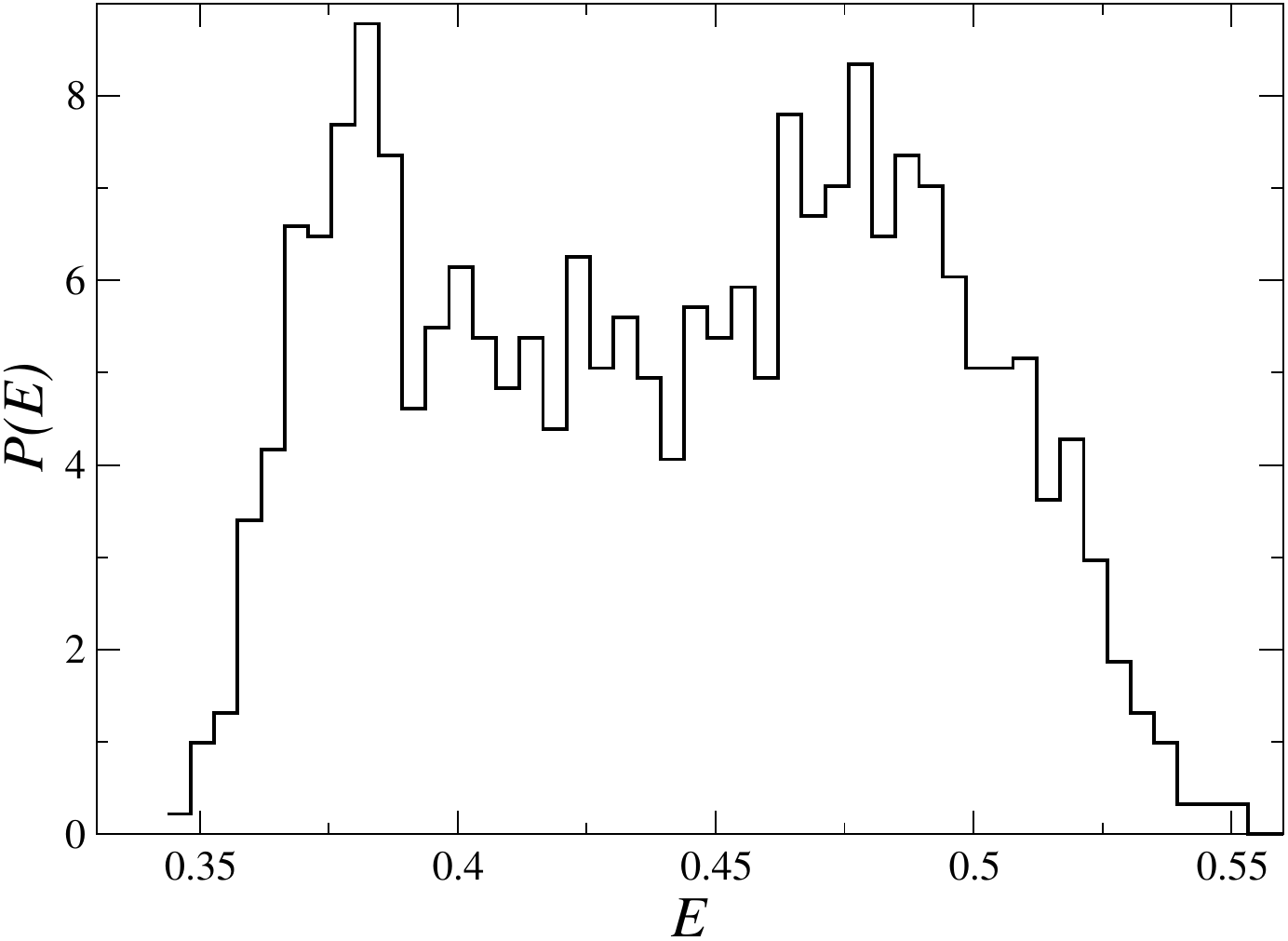}
  \caption{Top: Distribution of the energy $E$ for $v = 10$, $J=0.453$, $L=12$.
   Bottom: Distribution of the energy $E$ for $v = -10$, $J=0.3606$, $L=6$.
   Results for $N=6$ and $\kappa = 0$.}
\label{fig:histo-N6}
\end{figure}

For $N=6$ we expect first-order transitions for all values of $v$, For
$\kappa = 0$ and $N=7$, Ref.~\cite{PV-20-largeNCP} observed a very
strong metastability already on lattices of size $L=12$. Thus, we have
performed simulations on small lattices to be able to identify
metastability effects. We have considered two values of $v$, $v=10$
and $v=-10$.  In both cases we observe a bimodal distribution of the
energy in some interval of values of $J$. In Fig.~\ref{fig:histo-N6}
we show the probability distribution of
\begin{equation}
   E = {1\over 3 L^3} \sum_{{\bm x}\mu} 
      \bar{\bm z}_{\bm x} \cdot {\bm z}_{{\bm x} + \hat{\mu}} 
      \lambda_{{\bm x},\mu}, 
\end{equation}
for two specific values of $J$.
Data show a clear  two-peak structure with a large latent heat, confirming
the first-order nature of the transitions. 

\section{Conclusions} \label{sec4}

In this work we discuss the critical behavior of lattice Abelian gauge
models in which the fundamental field is an $N$-component complex
vector, and which are symmetric under $SO(N)$ transformations,
focusing on the behavior in the strong gauge-coupling regime.
A detailed analysis of the low-temperature configurations,
combined with general LGW arguments allowed Ref.~\cite{BPV-23c} to
make precise conjectures on the nature of the low-$\kappa$ transitions
in this class of models. In particular, while $SU(N)$ symmetric models
may undergo continuous transitions only for $N=2$ in the
strong-coupling regime, in $SO(N)$ symmetric models continuous
transitions (in the Heisenberg universality class) are also possible
for $N=3$, provided that the Hamiltonian parameters are such that the
symmetry breaking pattern at the transition is $SO(3)\to SO(2)\oplus
{\mathbb Z}_2$. For models with Hamiltonian~(\ref{LAH}) this occurs
for $v > 0$.

In this work we extend the theoretical analysis to values $N$
satisfying $N\ge 4$, focusing on the case $v > 0$, that was not
considered in Ref.~\cite{BPV-23c}. We perform a field-theoretical
analysis of the model, determining the RG flow of the renormalized
parameters close to four dimensions, using the $\epsilon$ expansion
approach. For $N\ge 5$ no stable fixed points are present, indicating
that the transitions in the strong-coupling regime must be always
of first order. For $N=4$, we can perform a nonperturbative analysis of
the RG flow, that allows us to prove the existence of a stable fixed
point, corresponding to  two decoupled Heisenberg critical behaviors.  Thus,
for $N=4$ continuous transitions are possible for $v > 0$, again in
the Heisenberg universality class.

The theoretical predictions of Ref.~\cite{BPV-23c} and those presented
here rely on several crucial assumptions. In particular, they assume
that an effective description can be obtained by considering the two
order parameters reported in Eq.~(\ref{wqoper}) and (\ref{wqopera})
($T^{ab}_L$ for $v > 0$ and $R^{ab}_L$ for $v < 0$), and the
corresponding LGW theory. To verify the correctness of these
assumptions, we have performed numerical simulations.  For $N=2$ we
observe an Ising transition and an XY transition for $v = 10$ and $v =
-10$. Heisenberg transitions are observed for $v=10$ both for $N=3$
and $N=4$---in the latter case with significant scaling corrections,
in agreement with theory, that predicts corrections decaying as
$L^{-0.19}$ with the size $L$ of the system. For $N=6$ transitions are
of first order for $v=10$ and $v=-10$.  The FSS analysis of the MC
data therefore fully confirms the general scenario.

\acknowledgments

The authors acknowledge support from project PRIN 2022 ``Emerging
gauge theories: critical properties and quantum dynamics''
(20227JZKWP).

\appendix 
\section{Mean-field analysis of the Landau-Ginzburg-Wilson model for an
antisymmetric tensor} \label{App}

We now determine the symmetry breaking patterns for the 
LGW theory with Lagrangian~(\ref{hlgt}). For this purpose it is enough to
consider the model in the mean-field approximation, i.e., to determine 
the minima of the mean-field Hamiltonian
\begin{eqnarray}
H_{MF} &=&  
  r \,\hbox{\rm Tr} \,\Psi^t\Psi +  u \, (\hbox{\rm Tr}
\,\Psi^t\Psi)^2 + w\, \hbox{\rm Tr}\, (\Psi^t\Psi)^2.\quad
\label{hlgtMF} 
\end{eqnarray}
As the Hamiltonian is $SO(N)$ invariant, we can use this symmetry to simplify 
the analysis. We will now show that every real antisymmetric matrix $A$ of 
rank $N$ can be written as $A = V A_B V^t$, where $V\in SO(N)$ and $A_B$ is a 
block-diagonal antisymmetric matrix.
If $N$ is even, we can write ($M=N/2$)
\begin{equation}
A_B = \hbox{diag} (A_1,\ldots,A_M),
\label{block1}
\end{equation}
where the matrices $A_i$ are antisymmetric and two-dimensional. If $N$ is 
odd, we have instead ($M = (N-1)/2$)
\begin{equation}
A_B = \hbox{diag} (A_1,\ldots,A_{M},0).
\label{block2}
\end{equation}
This result has been proved in Ref.~\cite{Youla-61} for complex matrices
($V$ is unitary in that case). Let us sketch here the derivation for real
matrices. Note that the nonvanishing eigenvalues of an antisymmetric matrix
are purely imaginary. Since $A$ is also real, they must appear in
complex-conjugate pairs.  Therefore, if $N$ is even the eigenvalues are
$\{ia_1,-ia_1,ia_2,-ia_2,\ldots\}$, where the coefficients $a_i$ are real. If
$N$ is odd one eigenvalue is necessarily zero. Since the matrix $A^t A = - A^2$
is symmetric, it can be diagonalized by using an orthogonal matrix. Therefore,
there exists an orthogonal matrix $V$ such that 
\begin{eqnarray}
&& \hbox{diag} (-a_1^2, -a_1^1, -a_2^2, -a_2^2, \ldots) = 
\nonumber \\
   && \qquad =  V A^2 V^t = (V A V^t) (V A V^t) .
\end{eqnarray}
Now consider an eigenvector $v$ of $A^2$. It is trivial to show that 
$Av$ is also an eigenvector of $A^2$ with the same eigenvalue. If all 
eigenvalues $a_i$ are distinct, this relation implies that 
$V A V^t$ has necessarily the block-diagonal structure (\ref{block1}) or 
(\ref{block2}). If not all eigenvalues are distinct, we can still choose 
$V$ so that the block structure holds.

It is interesting to note that a two-dimensional antisymmetric matrix has
the form 
\begin{equation}
\begin{pmatrix} 0 & a \\
               -a & 0 
\end{pmatrix}
\end{equation}
and thus it is determined by its eigenvalues $\pm i a$, up to a sign. 

We can now discuss the minima of the mean-field
Hamiltonian. If $M = \lfloor N/2\rfloor$, modulo $SO(N)$ 
transformations we can take $\Psi$ in block-diagonal form so that 
\begin{equation}
\Psi^t \Psi = \hbox{diag}(a_1^2, a_1^2, a_2^2, a_2^2, \ldots,
    a_M^2, a_M^2, (0) ),
\end{equation}
where the last 0 occurs only for odd $N$. We should therefore determine the 
minima of 
\begin{equation}
  H_{MF} =  2 r \sum_i a_i^2 + 4 u \Bigl(\sum_i a_i^2\Bigr)^2 +2 w \sum_i a_i^4.
\end{equation}
For $r > 0$, the minimum corresponds to $a_i = 0$ for all $i$: this is 
the disordered phase. For $r < 0$, we should distinguish two cases:

(i) For $w < 0$, a minimum configuration corresponds to $a_1 = a$,
$a_2,\ldots,a_M=0$, with
\begin{equation}
a^2 = -{r \over 2 (2 u + w)}, \quad H_{MF,\rm min} = - {r^2\over 2 (2
  u + w)} .
\end{equation}
The configuration is invariant under $SO(2)\oplus O(N-2)$
transformations (note that two-dimensional antisymmetric matrices are
invariant under $SO(2)$ transformations).  This is the relevant phase
for the model with $v > 0$.

(ii) For $w > 0$ the minimum corresponds to $a_1,\ldots,a_M=a$ with
\begin{equation}
a^2 = -{r \over 2 (2 M u + w)}, \quad 
H_{MF,\rm min} = - {M r^2\over 2 (2 M u + w)},
\end{equation}
which is invariant under the compact symplectic group USp($2M$).
If $N$ is odd there is an additional ${\mathbb Z}_2$ invariance.

Note that this calculation also provides the stability conditions for the 
quartic potential \cite{AKL-13}, $2 u + w> 0$ and $2 M u + w> 0$.

\end{document}